\documentclass[%
 aip,
 amsmath,amssymb,
 reprint,%
]{revtex4-1}

\usepackage{graphicx}
\usepackage{dcolumn}
\usepackage{bm}

\usepackage[utf8]{inputenc}
\usepackage[T1]{fontenc}
\usepackage{mathptmx}
\usepackage{etoolbox}

\makeatletter
\def\@email#1#2{%
 \endgroup
 \patchcmd{\titleblock@produce}
  {\frontmatter@RRAPformat}
  {\frontmatter@RRAPformat{\produce@RRAP{*#1\href{mailto:#2}{#2}}}\frontmatter@RRAPformat}
  {}{}
}%
\makeatother
\begin{document}

\preprint{AIP/123-QED}

\title[PSWS at millikelvin temperatures]{Propagating spin-wave spectroscopy in nanometer-thick YIG films at millikelvin temperatures}

    \author{Sebastian Knauer}
        \email{knauer.seb@gmail.com}
        \affiliation{University of Vienna, Faculty of Physics, A-1090 Vienna, Austria}
        \author{Krist\'{y}na Dav\'{i}dkov\'{a}}
        \affiliation{CEITEC BUT, Brno University of Technology, 612 00 Brno, Czech Republic}
    \author{David Schmoll}
        \affiliation{University of Vienna, Faculty of Physics, A-1090 Vienna, Austria}
        \affiliation{University of Vienna, Vienna Doctoral School in Physics, A-1090 Vienna, Austria}
    \author{Rostyslav O. Serha}
        \affiliation{University of Vienna, Faculty of Physics, A-1090 Vienna, Austria}
        \affiliation{University of Vienna, Vienna Doctoral School in Physics, A-1090 Vienna, Austria}
    \author{Andrey Voronov}
        \affiliation{University of Vienna, Faculty of Physics, A-1090 Vienna, Austria}
        \affiliation{University of Vienna, Vienna Doctoral School in Physics, A-1090 Vienna, Austria}
    \author{Qi Wang}
        \affiliation{University of Vienna, Faculty of Physics, A-1090 Vienna, Austria}
    \author{Roman Verba}
        \affiliation{Institute of Magnetism, Kyiv 03142, Ukraine}
    \author{Oleksandr V. Dobrovolskiy }
        \affiliation{University of Vienna, Faculty of Physics, A-1090 Vienna, Austria}
    \author{Morris Lindner}
        \affiliation{INNOVENT e.V. Technologieentwicklung, Prüssingstraße 27B, Jena, Germany}
    \author{Timmy Reimann}
        \affiliation{INNOVENT e.V. Technologieentwicklung, Prüssingstraße 27B, Jena, Germany}
    \author{Carsten Dubs}
        \affiliation{INNOVENT e.V. Technologieentwicklung, Prüssingstraße 27B, Jena, Germany}
    \author{Michal Urb\'{a}nek}
        \affiliation{CEITEC BUT, Brno University of Technology, 612 00 Brno, Czech Republic}
    \author{Andrii V. Chumak}
        \affiliation{University of Vienna, Faculty of Physics, A-1090 Vienna, Austria}

\date{\today}

\begin{abstract}
Performing propagating spin-wave spectroscopy of thin films at millikelvin temperatures is the next step towards the realisation of large-scale integrated magnonic circuits for quantum applications. 
Here we demonstrate spin-wave propagation in a $100\,\mathrm{nm}$-thick yttrium-iron-garnet film at the temperatures down to $45 \,\mathrm{mK}$, using stripline nanoantennas deposited on YIG surface for the electrical excitation and detection. The clear transmission characteristics over the distance of $10\,\mu \mathrm{m}$ are measured and the subtracted spin-wave group velocity and the YIG saturation magnetisation agree well with the theoretical values.
We show that the gadolinium-gallium-garnet substrate influences the spin-wave propagation characteristics only for the applied magnetic fields beyond $75\,\mathrm{mT}$, originating from a GGG magnetisation up to $47 \,\mathrm{kA/m}$ at $45 \,\mathrm{mK}$. 
Our results show that the developed fabrication and measurement methodologies enable the realisation of integrated magnonic quantum nanotechnologies at millikelvin temperatures.
\end{abstract}
\maketitle


\section{\label{sec:Introduction}Introduction}
Yttrium-Iron-Garnet (YIG, $\mathrm{Y}_{3}\mathrm{Fe}_{5}\mathrm{O}_{12}$) is the ideal choice of material to build and develop classical and novel quantum technologies~\cite{Barman2021, Chumak2022} by coupling spin waves, and their single quanta magnons, to phonons~\cite{Li2020}, fluxons~\cite{Dobrovolskiy2019}, or to microwave and optical photons~\cite{Tabuchi2016, Boventer2018, Bittencourt2019, Lachance-Quirion2019}. These technologies may be realised by the coupling to bulk or spherical YIG samples (e.g.~Ref.~\cite{Lachance-Quirion2020, Morris2017}), or by the fabrication of integrated structures in thin YIG films~\cite{Heinz2020, Wang2020}. Such nanometer-thick films can be grown using liquid phase epitaxy (LPE)~\cite{Dubs2017, Dubs2020}, exhibiting long spin wave propagation lengths, narrow linewidths and low damping constants~\cite{Henry1973, Hongxu1984, Dubs2017, Dubs2020}. Significant progress was made in realising YIG nano-waveguides with lateral dimensions down to $50\,\mathrm{nm}$~\cite{Heinz2020}, 
in understanding the spin-wave properties in these waveguides~\cite{Wang2019}, and in using them for room-temperature data processing~\cite{Wang2020}. 

To create, propagate and read out spin waves at single magnon level, millikelvin temperatures are required to suppress thermal magnons according to the Bose-Einstein statistics~\cite{Chumak2022}. The established technique of ferromagnetic resonance (FMR) spectroscopy was used to characterise YIG films of micrometer~\cite{Mihalceanu2018} and nanometer thicknesses~\cite{Jermain2017, Guo2022, Golovchanskiy2019} at kelvin temperatures. At millikelvin temperatures FMR measurements were performed on  micrometer~\cite{Kosen2019} and nanometer-thick~\cite{Jermain2017, Golovchanskiy2019, Guo2022} YIG films. Another method, propagating spin-wave spectroscopy (PSWS), is often used to characterise magnon transport between spatially-separated sources and detectors. This technique was successfully used in thin films at room~\cite{Vlaminck2010,Vanatka2021} and near room temperature~\cite{Alam2019}, and at millikelvin temperatures for micrometer-thick YIG slabs~\cite{VanLoo2018, SchmollDavid2022Etfh} and micrometer-scaled hybrid magnon-superconducting systems~\cite{Baity2021}.

The ability to process information in sub-$100\,\mathrm{nm}$ sized magnonic structures is one of the key advantages of magnonics, which translates also to the fields of hybrid opto-magnonic quantum systems and quantum magnonics. To couple PSW to these nanostructures efficiently at millikelvin temperatures, integrated nanoantennas~\cite{Yu2014,Vanatka2021} are required.
Here we demonstrate PSWS at millikelvin temperatures, with base temperatures reaching $45\,\mathrm{mK}$ in a $100\,\mathrm{nm}$-thick YIG film, using integrated nanoantennas separated by 10 micrometers for excitation and detection. The analysis is focused on magnetostatic surface spin waves (MSSWs, also called “Damon-Eshbach” mode) that propagate perpendicular to an in-plane magnetic field $\mathbf{k}\perp \mathbf{B} $. 
We find that magnon transport at the nanometer structure scale can be measured also down to millikelvin temperatures.
Although the propagation signal is measurable across a wide field and temperature range, we observe that the transmitted signal is distorted for applied magnetic fields above $75\,\mathrm{mT}$. This effect is largely caused by the magnetisation of the gadolinium-gallium-garnet (GGG) substrate, on which the YIG film is grown. It reaches $47\,\mathrm{kA/m}$ for  $75\,\mathrm{mT}$ of applied external magnetic field at $45\,\mathrm{mK}$ temperature. In general, our findings agree with the increase in the damping of YIG grown on GGG at low-temperatures reported in the literature~\cite{Jermain2017, Kosen2019, Golovchanskiy2019, Guo2022}. 

\begin{figure*}[h!t]
    \centering
    \includegraphics[width=0.99\textwidth]{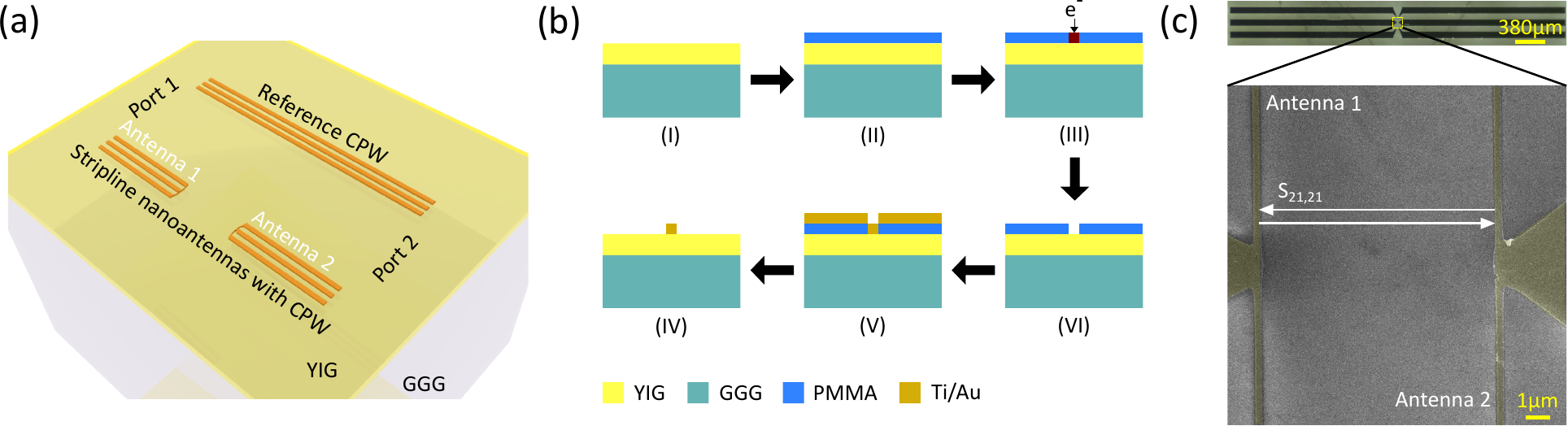}
    \caption{
    {\bf Overview of the electron-beam lithographed stripline nanoantennas on the yttrium-iron-garnet film.
    }
    (a) Sketch of the sample used in these measurements. Stripline nanoantennas coupled to coplanar waveguide (CPW) and the reference CPW are fabricated atop a $100\,\mathrm{nm}$-thick yttrium-iron-garnet film on a $500\,\mathrm{\mu m}$-thick gadolinium-gallium-garnet substrate.
    (b) The coplanar-waveguide coupled nanoantennas are fabricated with electron-beam lithography. These nanoantennas are made of Ti($5\,\mathrm{nm}$)/Au($55\,\mathrm{nm}$) (more details in main text).
    (c) Optical and secondary-electron images of the CWP nanoantennas used in the manuscript. These nanoantennas are $10\,\mathrm{\mu m}$ spaced apart and have a width of $330\,\mathrm{nm}$ and length of $120\,\mathrm{\mu m}$. The propagating spin waves (PSW) are excited and detected by the stripline nanoantennas 1 and 2 respectively. The transmission is measured through the S-parameters, acquired by a vector network analyser.
    }
    \label{fig:Fig1}
\end{figure*}
First, we explain the sample preparation and experimental techniques, before we continue to pre-characterise the sample at room and base temperature, using standard FMR techniques. Then we discuss the first PSWS experiment, in which a fixed external magnetic field is applied and the temperature is swept from base to room temperature. We continue to analyse the propagation characteristics in more detail by comparing the low-temperature measurements to room-temperature results and extract the spin-wave group velocities. Finally, we perform PSWS at higher external magnetic fields, to investigate the propagation characteristics between the room and base temperature. The magnetisation of GGG is measured by vibrating sample magnetometry (VSM) of a GGG-only substrate at low-temperatures.

\section{\label{sec:Sample}Sample and experimental setups}
In our experiments we use an LPE-grown $100\,\mathrm{nm}$-thick $(111)$-orientated YIG film on a $500\,\mathrm{\mu m}$-thick GGG substrate, as sketched in Fig.~\ref{fig:Fig1}~(a). Atop the YIG film we fabricate nanoantennas connected to CPWs, using an electron-beam lithography process Fig.~\ref{fig:Fig1}~(b). 
First, a single layer of PMMA is spin-coated and baked. After, we use electron-beam lithography to write the antenna structures, develop the sample and deposit a layer of  Ti($5\,\mathrm{nm}$)/Au($55\,\mathrm{nm}$), using electron beam physical vapour deposition, followed by lift-off. Figure~\ref{fig:Fig1}(c) shows an optical (top) and a secondary-electron image (bottom) with the coplanar waveguides and stripline nanoantennas used in this work. Here the nanoantennas have a spacing of $10\,\mathrm{\mu m}$. The stripline nanoantennas possess a width of $330\,\mathrm{nm}$ and a length of $120\,\mathrm{\mu m}$. Additionally, we fabricate a reference coplanar waveguide, to measure the FMR signals only (see Fig.~\ref{fig:Fig1}~(a)).
After fabrication, the sample is glued and then wire-bonded, with a $75\,\mathrm{\mu m}$ diameter gold wire, to a high-frequency printed circuit board and mounted into the dilution refrigerator.

Our setup is based on a cryogenic-free dilution refrigerator system (BlueFors-LD250), which reaches base temperatures below $10\,\mathrm{mK}$ at the mixing chamber stage. The sample space possesses a base temperature of about $16\,\mathrm{mK}$. During operation, the sample space heats up to about $45\,\mathrm{mK}$. At these temperatures, the thermal excitations of gigahertz-frequency magnons and phonons are still suppressed. The input signal is transmitted and collected from the sample (ports 1 and 2 Fig.~\ref{fig:Fig1}(a)), using high-frequency copper and superconducting wiring  each attenuated by $7\,\mathrm{dB}$ to reduce thermal noise.
The signals are collected with a $70\,\mathrm{GHz}$ vector network analyser (Anritsu VectorStar MS4647B).

The room-temperature measurements are carried out on a home-built setup. The setup consists of a VNA (Anritsu MS4642B) connected to an H-frame electromagnet GMW 3473-70 with an $8\,\mathrm{cm}$ air gap for various measurement configurations and magnet poles of $15\,\mathrm{cm}$ diameter to induce a sufficiently uniform biasing magnetic field. The electromagnet is powered by a bipolar power supply BPS-85-70EC (ICEO), allowing it to generate up to $0.9\,\mathrm{T}$ at $8\,\mathrm{cm}$ air gap. The input powers are adjusted, to obtain the same power levels at the sample as in the cryogenic measurements, to account for cable losses and the previously mentioned attenuators. The precise microwave powers for each individual experiment are stated later.
\section{\label{sec:Results}Results and Discussion}
\begin{figure}[h!t]
    \centering
    \includegraphics[scale=0.39]{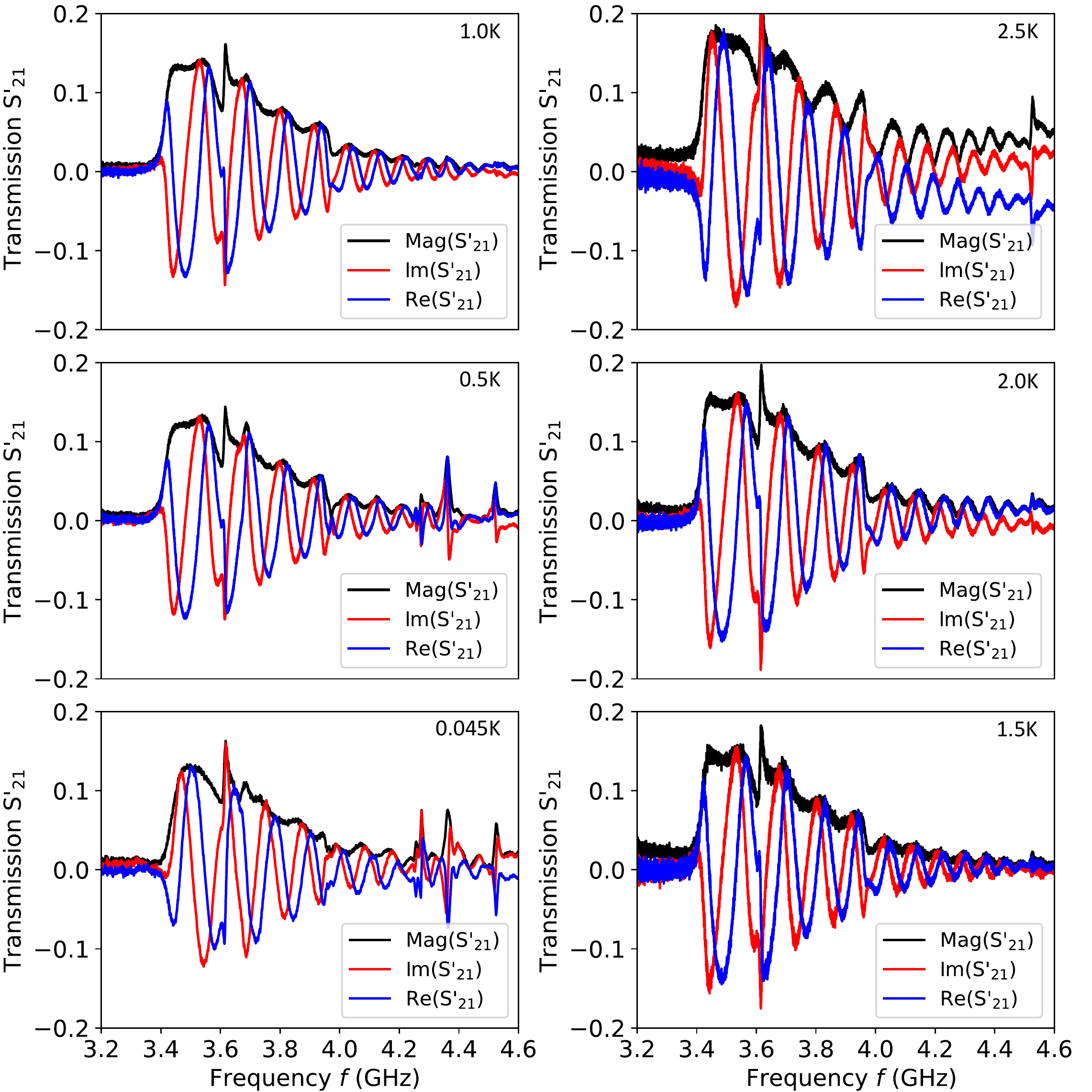}
    \caption{
   {\bf Linear magnitude, real and imaginary part of the $\mathrm{S'_{21}}$ parameters for propagating spin waves (PSW) in the Damon-Eshbach mode, using $\mathbf{50\,\mathrm{\textbf{mT}}}$ of external magnetic field and different temperatures.} The applied microwave power was set to $-28\,\mathrm{dBm}$ (at the sample) with an average sampling of 50 for $45\,\mathrm{mK}$-$1\,\mathrm{K}$ and 100 for $1.5\,\mathrm{K}$-$2.5\,\mathrm{K}$. The FMR point ($k=0$) is constant at $3.36\,\mathrm{GHz}$ ($189\,\mathrm{kA/m}$) for all measured PSW.
   }
    \label{fig:Fig2}
\end{figure}
First, we use the reference CPW to pre-characterise the sample and to estimate the Gilbert damping $\alpha_{\mathrm{s}}$. 
We plot our FMR and PSWS data according to
\begin{eqnarray}
S'_{\mathrm{21}}(f) = \frac{S_{\mathrm{21,sig}}(f) - S_{\mathrm{21,ref}}(f)}{S_{\mathrm{21,ref}}(f)},
\label{eq:PSWS_Sparameter}
\end{eqnarray}
where $f$ denotes the set of frequency points of the complex transmission  signal $S_{\mathrm{21,sig}}(f)$ and reference $S_{\mathrm{21,ref}}(f)$ values~\cite{Kalarickal2006}. The reference signal is obtained by detuning the external magnetic field by $+50\,\mathrm{mT}$.
For the FMR reference measurements, we find the Gilbert damping parameter $\alpha_{\mathrm{s}} = (5.98\pm 0.3)\times 10^{-4}$ for room temperature, and $\alpha_{\mathrm{s}} = (3\pm 1.5)\times 10^{-3}$ at $45\,\mathrm{mK}$ respectively. The large error in the low-temperature case originates from the fit uncertainty in the slope of FMR linewidth versus FMR frequency. The methodology developed in Ref.~\cite{Kalarickal2006}, which accounts for asymmetry and phase offset in the distorted FMR signal, was used.
The order of magnitude in the Gilbert damping at $45\,\mathrm{mK}$ is in good agreement with previously reported values for thin YIG films at Kelvin temperatures~\cite{Jermain2017}. 

We perform the first PSWS experiment at a fixed external magnetic field of $50\,\mathrm{mT}$, using the stripline nanoantennas shown in Fig.~\ref{fig:Fig1}. Figure~\ref{fig:Fig2} displays the linear magnitude (black), real (blue) and imaginary (red) part of the transmission data (cw-mode), together with a temperature sweep from the base temperature of $45\,\mathrm{mK}$ up to $2.5\,\mathrm{K}$, i.e.~about the Curie-Weiss temperature of GGG~\cite{Petrenko1999, Sabbaghi2020}. The spin waves are excited with a power of $-28\,\mathrm{dBm}$ at the sample, with the external magnetic field applied perpendicular to the propagation direction. In Fig.~\ref{fig:Fig2} we verify the ability to measure the transmission across the entire temperature range and observe a propagation signal with a fixed FMR point ($k=0$) of about $3.36\,\mathrm{GHz}$, corresponding to an effective saturation magnetisation of about $189\,\mathrm{kA/m}$. The signal amplitude increases by about $30\%$ from $45\,\mathrm{mK}$ to $2.5\,\mathrm{K}$.

\begin{figure}[t!]
    \centering
    \includegraphics[width=0.49\textwidth]{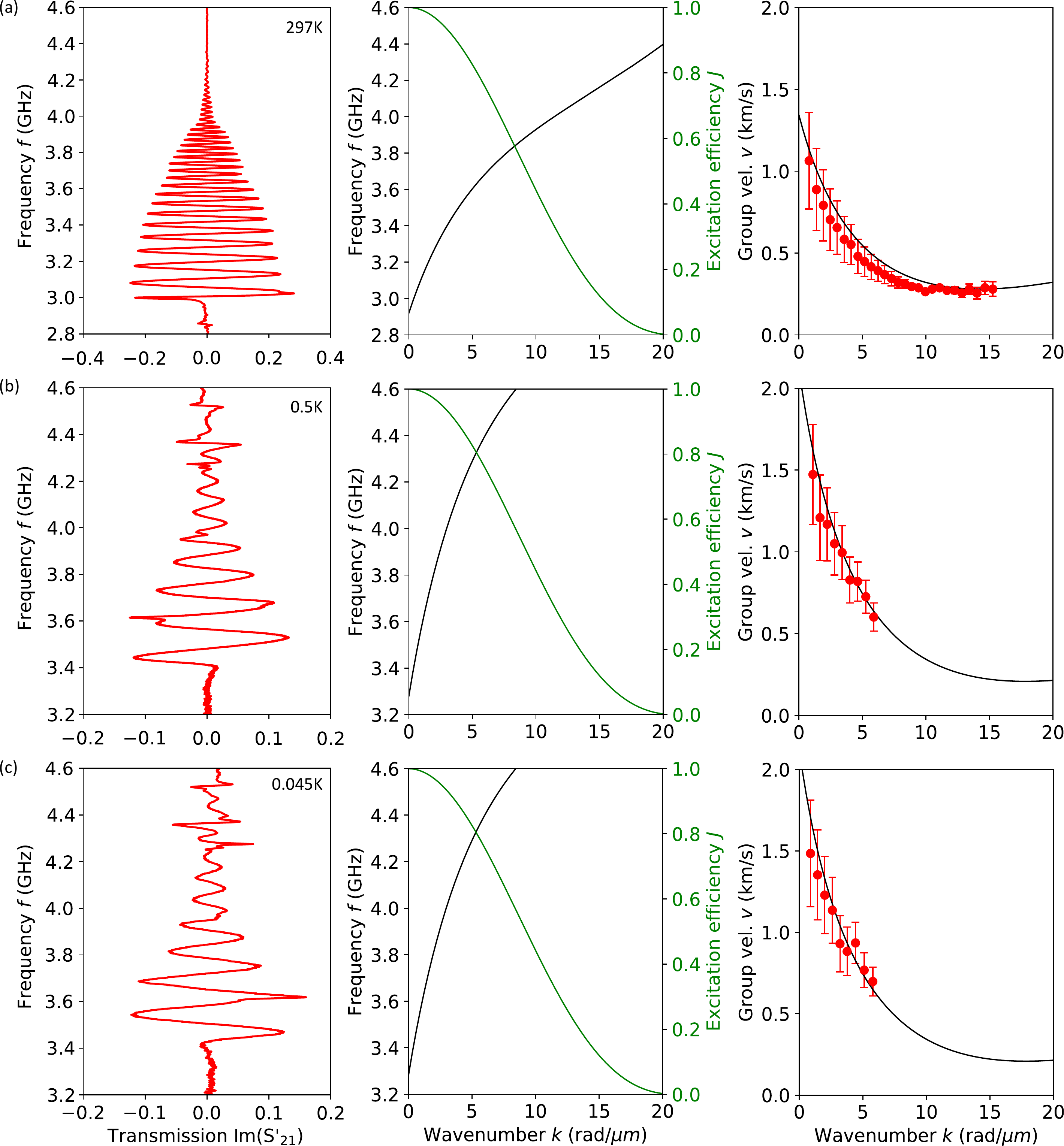}
    \caption{
    {\bf Imaginary part of the $\mathrm{S'_{21}}$ parameter, calculated dispersion relation, antenna excitation efficiency and group velocity for PSW (Damon-Eshbach mode), using $\mathbf{50\,\mathrm{\textbf{mT}}}$ external field at different temperatures.} 
    The theoretical group velocity is calculated as the derivation of the dispersion relation and measured as $v_g={\delta}f \cdot D$, with the periodicity of the transmission in the Im($\mathrm{S'_{21}}$) parameters ${\delta}f$ and the gap between the nanoantennas $D$ (see Ref.~\cite{Vlaminck2010}). The parameters measured and used for the calculation are the following:
    (a) $297\,\mathrm{K}$, $M_s=142\,\mathrm{kA/m}$,
    (b) $500\,\mathrm{mK}$, $M_s=189\,\mathrm{kA/m}$,
    (c) $45\,\mathrm{mK}$, $M_s=189\,\mathrm{kA/m}$.
    The effective saturation magnetisation increases and thus group velocity increases by about $50\,\%$ at millikelvin temperatures.
    } 
    \label{fig:Fig3}
\end{figure}

\begin{figure*}[t!h]
    \centering
    \includegraphics[width=0.99\textwidth]{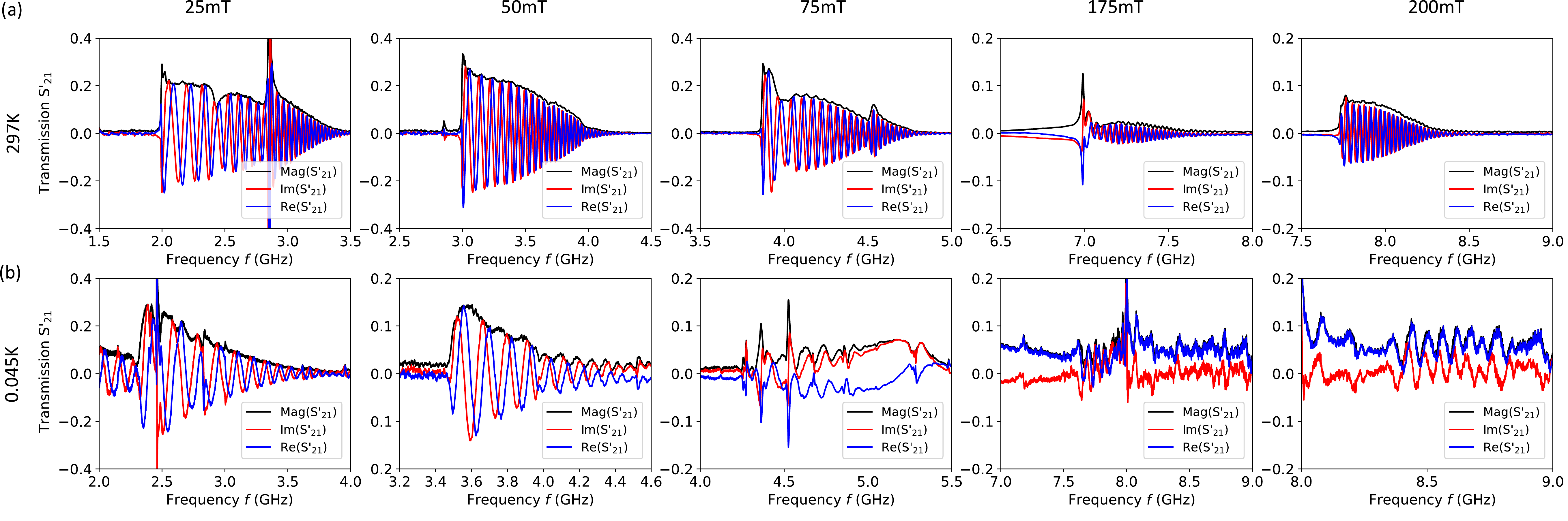}
    \caption{
    {\bf 
    Linear magnitude, real and imaginary part of the $\mathrm{S'_{21}}$ parameters for PSWS in the Damon-Eshbach mode at different external fields. The applied microwave power was set to $\mathbf{-28\,\mathrm{\textbf{dBm}}}$ (at the sample) with an averaging of 10 (for $\mathbf{297K}$) and 25 (for $\mathbf{0.045K}$).
    }\newline
    (a) Room temperature ($297\,\mathrm{K}$): The spin-wave propagation can be measured over a wide magnetic field range. 
    (b) Base temperature ($45\,\mathrm{mK}$): The spin-wave propagation for magnetic fields in the range from about $25\,\mathrm{mT}$ to $75\,\mathrm{mT}$ is trackable, while above $75\,\mathrm{mT}$ the magnitude and its propagation characteristics start to be distorted. This effect is a result of the increased magnetisation of the GGG substrate (see Fig.~\ref{fig:Fig6}). 
    } 
    \label{fig:Fig4}
\end{figure*}

We continue to investigate the spin-wave propagation in more detail and compare the results to room-temperature measurements. Figure~\ref{fig:Fig3} (first column) depicts the imaginary part of the $\mathrm{S'_{21}}$ parameters for PSWs between the two nanoantennas at three different selected temperatures: (a) $297\,\mathrm{K}$, (b) $500\,\mathrm{mK}$, and (c) $45\,\mathrm{mK}$, at a fixed external magnetic field of $50\,\mathrm{mT}$. The second column in Fig.~\ref{fig:Fig3} shows the corresponding calculated dispersion relations for MSSWs (black), using the Kalinikos-Salvin model~\cite{Kalinikos1986}. The maximum excitation efficiency $J$ (green line Fig.~\ref{fig:Fig3}) is governed by the $330\,\mathrm{nm}$ stripline nanoantennas~\cite{Vlaminck2010}. 
The third column in Fig.~\ref{fig:Fig3} shows the theoretical group velocities as the derivation of the dispersion relation (black curve) and the measured values given by $v_g={\delta}f \cdot D$ (red dots), where ${\delta}f$ is the periodicity of the oscillations in the Im$(\mathrm{S'_{21}})$ parameters and $D$ the gap between the nanoantennas~\cite{Vlaminck2010}. The errors in the calculated group velocities are estimated from the error propagation of the frequency reading. %
We observe a reduction in propagation amplitude by about $50\%$ between the room and both cryogenic temperatures caused by the increase in Gilbert damping. We find values for the effective saturation magnetisation of $M_s=142\,\mathrm{kA/m}$ at room temperature and $M_s=189\,\mathrm{kA/m}$ for $45\,\mathrm{mK}$ and $500\,\mathrm{mK}$. The constant effective saturation magnetisation at millikelvin temperatures is in good agreement with literature~\cite{Hansen1974,Popov2011book},
with a value close to the observed ones in micrometer-thick YIG samples~\cite{VanLoo2018}.
In accordance with the increase in effective saturation magnetisation, we observe an increase of the group velocity by about $50\,\%$. The measured values are in good agreement with the theoretically calculated group velocities.

We continue our investigations by comparing the spin-wave propagation for higher external magnetic fields than in the previous measurements, at $297\,\mathrm{K}$ (Fig.~\ref{fig:Fig4}~(a)) and $45\,\mathrm{mK}$ (Fig.~\ref{fig:Fig4}~(b)). Figure~\ref{fig:Fig4} shows the linear magnitude (black), real (blue) and imaginary (red) part for PSW in the Damon-Eshbach mode at selected magnetic fields.
At room temperature, we measure the spin-wave signal over a wide external magnetic field range up to about $900\,\mathrm{mT}$. Examples for low fields are given in Fig.~\ref{fig:Fig4}~(a). However, at $45\,\mathrm{mK}$ the propagation characteristics are changing (Fig.~\ref{fig:Fig4}~(b)). After about $75\,\mathrm{mT}$ the magnitude of the spin-wave signal is reduced significantly and only a signature in the oscillation behaviour can be observed. Moreover, the fixed phase relation between the imaginary and real parts disappears, causing challenges in plotting the linear magnitude of the propagation signal. Examples for the reduced spin-waves signals are given for $175\,\mathrm{mT}$ and $200\,\mathrm{mT}$. This opposing behaviour between the room and base temperature is a clear indication, that beyond an external field of about $75\,\mathrm{mT}$ the GGG substrate magnetises enough to influence the propagation characteristics of the spin waves. Thus, future millikelvin measurements at high magnetic fields may rely on suspended YIG membranes or triangular nanostructures, which have already been demonstrated in other material systems (e.g. Ref.~\cite{Burek2014}).

\begin{figure}[t!]
    \centering
    \includegraphics[scale=0.6]{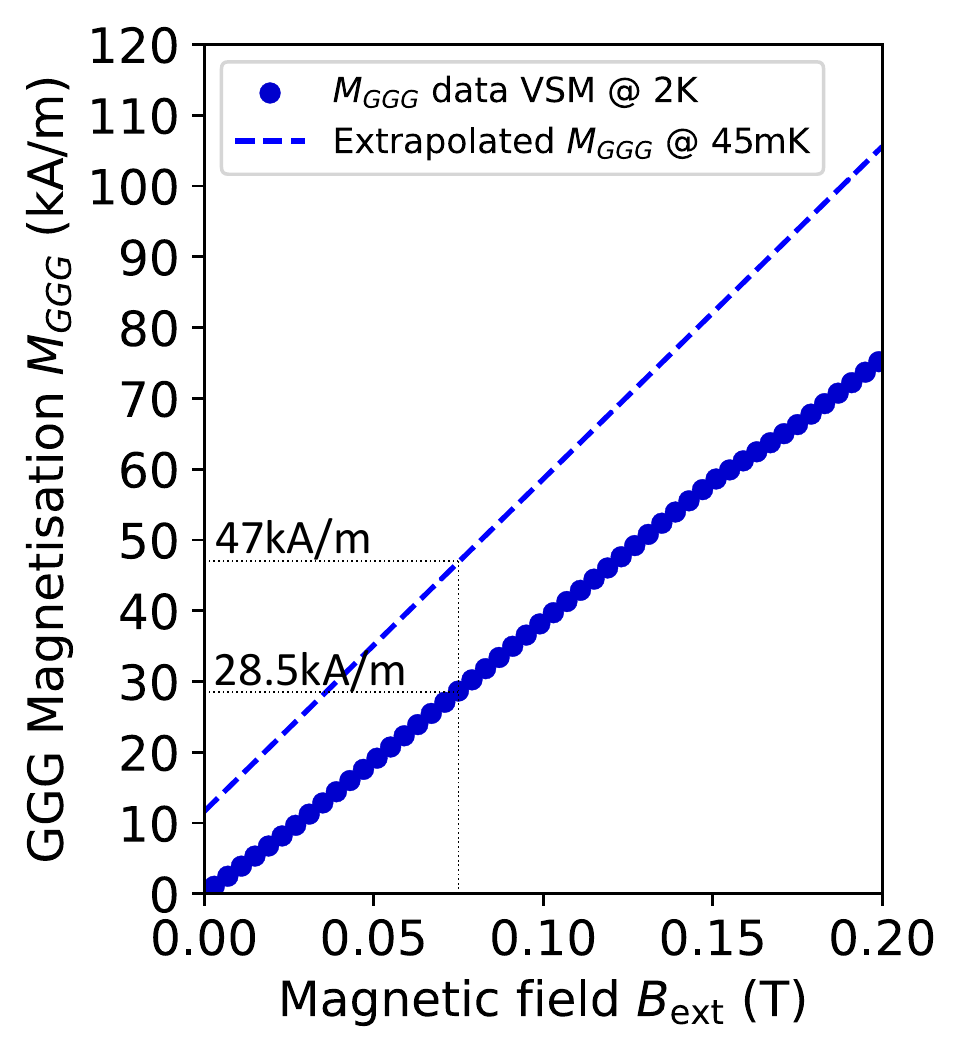} 
    \caption{
    {\bf Magnetisation of the GGG substrate versus the applied magnetic field.} A GGG-only sample is measured using a vibrating sample magnetometer (VSM) at $2\,\mathrm{K}$ (dark-blue dots), leading for example to an effective magnetisation of $28.5 \,\mathrm{kA/m}$. From the data the magnetisation values for $45\,\mathrm{mK}$ are extrapolated (blue dashed line). At $75\,\mathrm{mT}$ the magnetisation increases to about $47 \,\mathrm{kA/m}$.
    } 
    \label{fig:Fig6}
\end{figure}
To estimate the influence of the paramagnetic GGG substrate on the spin-wave propagation in YIG, we conclude our investigations by measuring the GGG magnetisation $M_{\mathrm{GGG}}$ of a $4\times4\times0.5\,\mathrm{mm}$ GGG-only substrate, using a vibrating sample magnetometer (VSM) in the temperature range from $2\,\mathrm{K}$ to $300\,\mathrm{K}$ in the presence of fields up to $9\,\mathrm{T}$. The results at $2\,\mathrm{K}$ for our magnetic fields of interest are shown in Fig.~\ref{fig:Fig6} (dark-blue dots). As the VSM is limited to kelvin temperatures, we extrapolate magnetisation values for GGG at $45\,\mathrm{mK}$ (blue dashed line), using the $2\,\mathrm{K}$ data. For example at $75\,\mathrm{mT}$ (Fig.~\ref{fig:Fig6} black dots) we find, that GGG possesses a magnetisation value of $28.5 \,\mathrm{kA/m}$ at $2\,\mathrm{K}$, which increases to about $47 \,\mathrm{kA/m}$ at $45\,\mathrm{mK}$. Thus, the temperature and magnetic field dependant GGG magnetisation may explain the observed reduction in the PSW amplitudes and the propagation distortions above external magnetic fields of $75\,\mathrm{mT}$.

However, the role of the paramagnetic GGG substrate on spin waves in YIG is the subject of separate systematic studies. Our PSWS measurements, supported by the FMR and VSM studies, suggest that the magnetic moment induced in GGG at millikelvin temperatures by the application of relatively large magnetic fields is at least partly responsible for the increase in spin-wave damping. The increase in the Gilbert damping constant $\alpha$ can only be approximately quantified, as this requires plotting the FMR linewidth $\Delta B$ against the FMR frequency $f_\mathrm{FMR}$ over a wide range of applied fields. However, since the FMR linewidth depends on the degree of the magnetisation of the GGG (given by the temperature and the applied field - see Fig.~\ref{fig:Fig6}), the dependence $\Delta B(f_\mathrm{FMR})$ becomes nonlinear and the parameter $\alpha$ loses its original physical meaning. Moreover, the measurement of FMR on nanometer-thick samples requires the careful subtraction of the reference microwave transmission signal (see Eq.~\ref{eq:PSWS_Sparameter}) at a $50\,\mathrm{mT}$ detuned magnetic field. Since this reference signal also depends significantly on the GGG magnetisation at low-temperatures, the measurement uncertainties increase. Nevertheless, we can qualitatively conclude that the increase in spin-wave damping in the nanometer-thick YIG films on GGG corresponds to the previously reported increase in damping in the micrometer-thick films on GGG~\cite{Mihalceanu2018, Jermain2017, Kosen2019}. Other phenomena that could contribute to the distortion of the PSWS experiments at the nanoscale at fields above $75\,\mathrm{mT}$ are the possible dependence of the magnetocrystalline anisotropy caused by the dependence of the YIG/GGG lattice mismatch on temperature and the absorption/distortion of the microwave signal in the CPW transmission lines (see Fig.~\ref{fig:Fig1}(a)) by the magnetised GGG substrate.


\section{\label{sec:Conclusions}Conclusions}
In conclusion, we have shown for the first time that propagating spin-wave spectroscopy in $100\,\mathrm{nm}$-thin YIG films can be performed in a wide temperature range, from millikelvin to room temperature, without changing the propagation characteristics.
At a fixed external magnetic field of $50\,\mathrm{mT}$ we confirm that the propagating spin waves maintain a constant ferromagnetic resonance frequency below temperatures of about $2.5\,\mathrm{K}$. However, the signal amplitude increases by $30\%$ between $45\,\mathrm{mK}$ and $2.5\,\mathrm{K}$, and further by about $50\%$ when the temperature is raised to room temperature.
In contrast to previous work we demonstrate, that only beyond an external field of about $75\,\mathrm{mT}$ the GGG substrate magnetises up to $47 \,\mathrm{kA/m}$ influence the spin-wave propagation at low-temperatures. 
With our experiments, we illustrate that although the GGG substrate influences the spin-wave propagation characteristics at millikelvin temperatures, future large-scale integrated YIG nanocircuits can be realised and measured.


\begin{acknowledgments}
The authors thank Vincent Vlaminck for useful discussions and feedback. SK acknowledges the support by the H2020-MSCA-IF under the grant number 101025758 (OMNI). KD was supported by the Erasmus+ program of the European Union. The authors acknowledge CzechNanoLab Research Infrastructure supported by MEYS CR (LM2018110).
The work of CD was supported by the Deutsche
Forschungsgemeinschaft (DFG, German Research Foundation) under grant 271741898. The work of ML was supported by the German Bundesministerium für Wirtschaft und Energie (BMWi) under grant 49MF180119. CD  thanks O. Surzhenko and R. Meyer (INNOVENT) for their support. The authors thank Oleksandr Dobrovolskiy for his support in the initial configuration of the dilution refrigerator.
\end{acknowledgments}

\section*{Author Declarations}
\subsection*{Conflict of Interest}
The authors have no conflicts to disclose.

\subsection*{Authors Contributions}
SK and MU conceived the experiment in discussion with AC.
SK and KD performed the experiments under the guidance of MU and AC.
SK and KD analysed and interpreted the data with support from AC.
RS and OD performed the VSM measurements at kelvin temperatures, and AV interpolated the data for millikelvin temperatures.
ML and TR prepared the LPE sample. CD conceived and supervised the LPE film growth.
QW and RV supported the measurements with theoretical expertise.
DS and SK set up the cryogenic system.
RS supported the measurements and analysis of the measurements.
SK wrote the manuscript with support from all co-authors.

\section*{Data Availability}
The data that support the findings of this study are available from the corresponding author upon reasonable request.

\section*{References}
\nocite{*}
\bibliography{aipsamp}


\end{document}